\documentclass[%
reprint,
superscriptaddress,
 amsmath,amssymb,
 aps,
 pra,
]{revtex4-2}

\usepackage[T1]{fontenc}
\usepackage[english]{babel}
\usepackage{siunitx}
\usepackage[utf8]{inputenc}
\usepackage{url}
\usepackage{dsfont}
\usepackage{bbm}
\usepackage{nicefrac}
\usepackage{setspace}
\usepackage[short]{optidef}
\usepackage{float}
\usepackage{comment}
\usepackage{dsfont}
\usepackage{amssymb}  
\usepackage[colorlinks=true, citecolor=blue,urlcolor=blue,linkcolor=blue,filecolor=black]{hyperref}
\usepackage[colorinlistoftodos, color=green!40, prependcaption]{todonotes}
\usepackage[normalem]{ulem}
\usepackage{amsthm}
\usepackage{mathtools}
\usepackage{physics}
\usepackage{graphicx}
\usepackage{adjustbox}
\usepackage{placeins}
\usepackage{lipsum}
\usepackage{csquotes}
\usepackage{soul}
\usepackage{graphicx}% Include figure files
\usepackage{dcolumn}% Align table columns on decimal point
\usepackage{bm}% bold math
\usepackage{listings}

\begin{document}

\preprint{AIP/123-QED}

\title[]{
Single-Photon-Based Clock Analysis and Recovery in Quantum Key Distribution}

\author{M. Zahidy}
%\email{muzah@fotonik.dtu.dk}
\affiliation{ 
Centre for Silicon Photonics for Optical Communications (SPOC), Department of Electrical and Photonics Engineering, Technical University of Denmark, Kgs. Lyngby, Denmark.
}

\author{D. Ribezzo}
\affiliation{Istituto Nazionale di Ottica del Consiglio Nazionale delle Ricerche (CNR-INO), 50125 Firenze, Italy}
\affiliation{Università degli Studi di Napoli Federico II, Napoli, Italy}

\author{R. M\"{u}ller}
\affiliation{ 
Centre for Silicon Photonics for Optical Communications (SPOC), Department of Electrical and Photonics Engineering, Technical University of Denmark, Kgs. Lyngby, Denmark.
}

\author{J. Riebesehl}
\affiliation{ 
Centre for Silicon Photonics for Optical Communications (SPOC), Department of Electrical and Photonics Engineering, Technical University of Denmark, Kgs. Lyngby, Denmark.
}

\author{A. Zavatta}
\affiliation{Istituto Nazionale di Ottica del Consiglio Nazionale delle Ricerche (CNR-INO), 50125 Firenze, Italy}
\affiliation{QTI S.r.l.,  50125, Firenze, Italy}

\author{M. Galili}
\affiliation{ 
Centre for Silicon Photonics for Optical Communications (SPOC), Department of Electrical and Photonics Engineering, Technical University of Denmark, Kgs. Lyngby, Denmark. 
}

\author{L. K. Oxenløwe}
\affiliation{ 
Centre for Silicon Photonics for Optical Communications (SPOC), Department of Electrical and Photonics Engineering, Technical University of Denmark, Kgs. Lyngby, Denmark. 
}

\author{D. Bacco}
\email{davide.bacco@unifi.it}
\affiliation{QTI S.r.l.,  50125, Firenze, Italy}
\affiliation{ 
Department of Physics and Astronomy,
University of Florence, 50019, Firenze, Italy
}

\begin{abstract}
Quantum key distribution is one of the first quantum technologies ready for the market. Current quantum telecommunication systems usually utilize a service channel for synchronizing the transmitter (Alice) and the receiver (Bob). However, the possibility of removing this service channel and exploiting a clock recovery method is intriguing for future implementation, both in fiber and free-space links. In this paper, we investigate criteria to recover the clock in a quantum communication scenario, and experimentally demonstrated the possibility of using a quantum-based clock recovery system in a time-bin quantum key distribution protocol. The performance of the clock recovery technique, in terms of quantum bit error rate and secret key rate, is equivalent to using the service channel for clock sharing.
\end{abstract}

\maketitle

\section{Introduction}
\label{Sec::Intro}
In recent years, quantum key distribution (QKD) has become a mature technology and is recognized by potential users and customers worldwide\cite{chen_2021,dynes_2019,wengerowsky2019,yin_2020,florence_field_trial, indu, euroqci_2}. QKD is a method to distribute symmetric keys exploiting the fundamental laws of quantum physics and can guarantee unconditionally secure cryptographic keys\cite{RevModPhys.81.1301,Pirandola:20}. During the last 30 years, the research community has investigated the QKD field from many perspectives and has achieved impressive results with regard to the channel length, secret key rate, implementation, multiplexing, encoding scheme, and type of source\cite{Stucki_2009,Pittaluga2021,Liao2017,vagniluca2020efficient,Zahidy_2021_OAM,murtaza2022efficient}.
The next step is integrating the QKD equipment with classical telecommunication networks and establishing quantum networks between cities and countries. In this direction, China has been the first country to realize a metropolitan area quantum network (MAN) for commercial applications\cite{Chen2021_46node}, and Europe has launched a concrete plan to create a European quantum communication infrastructure\cite{euroqci_2}. In this regard, some efforts to establish an inter-European QKD network have already been achieved\cite{Ribezzo_3cities}.
Concerning the implementation of the quantum networks, there are important requirements in terms of form factors, number of additional service channels, number of users, energy consumption, etc.,  which make integrating QKD systems with our current communication infrastructure easier. For example, most of the current commercial quantum key distribution systems use a service channel for distributing a clock synchronization signal between the transmitter (Alice) and the receiver (Bob)\cite{Chen2021_46node,Ribezzo_3cities}. In communication, clock synchronization is necessary to generate the correct stream of data for post-processing. In order to recover the transmitted data, it is expected that detection time at the receiver is in accordance with the transmitter, up to a fixed offset, which corresponds to the time of flight of the signal. In the absence of a true non-drifting clock (atomic clock), the recorded time of detections at the receiver varies with respect to the transmitter. This effect leads to loss of periodicity and incorrect decoding of data. In such a scenario, clock synchronization is necessary to prevent data corruption.

From the perspective of resource optimization, overcoming the need for clock synchronization would introduce multiple benefits. Locking to a reference clock and transmitting a synchronization signal\cite{Sasaki:11} are the most common methods of sharing a clock. In the former, a reference clock such as a rubidium reference clock\cite{Steinlechner2017,FSQKD_Marcikic} or global navigation satellite system disciplined oscillators\cite{PhysRevA.92.052339,Avesani2021} provides a reference clock while in the latter, a synchronization laser together with an optical clock recovery instrument is used. Both methods rely on auxiliary pieces of equipment, with the latter occupying an optical communication channel as well. Co-propagation of quantum and clock signals with wavelength multiplexing can introduce issues with background noise (i.e., Raman, Brillouin, and Rayleigh scattering) due to limited extinction ratio in filtering and additional losses due to filtering, which degrades the key generation rate. Furthermore, reserving a separate fiber link, which is a valuable resource, and necessary amplification of the signal at long distances add to the operational cost of QKD systems.

Exploiting transmitted qubits to recover the clock and extract timing information is proposed and investigated on multiple occasions \cite{PhysRevApplied.13.054041,QubitBasedSynch_Cochran,Takenaka2017,Wang:21}, as well as proposals for timing synchronization that requires an entangled source\cite{Shi2022} or a photon pair\cite{10.1063/1.5121412,10.1063/5.0031166,10.1063/1.5086493}. 
The success of a clock recovery depends on multiple factors and a thorough study of such factors is lacking. A hardware malfunction or processing limitations result in uncertainties that make the task of clock recovery costly, if not impossible.

In this paper, we discuss some of the criteria for a successful clock recovery and introduce a technique to remove the errors that emerge in a clock recovery process. The technique is based on the fast Fourier transform (FFT) of small blocks of data acquired directly from the transmitted qubits of the QKD system. Additionally, we present a novel optimization method to further increase the precision of the clock recovery, and simulate multiple realizations to show its performance. We tested the introduced technique in a synchronization-free quantum key distribution employing a time-bin encoding scheme with one decoy state, which is the most common in commercial QKD systems. Furthermore, as a case study, we demonstrated that the clock recovery solution can be employed for more than 30 dB of channel loss - for given detectors' performance and receiver's loss - and the performance in terms of quantum bit error rate (QBER) and secret key rate (SKR) is equivalent to a clock-assisted QKD system. Our solution can pave the way for easier integration of QKD systems into the existing telecom networks.

\section{Method}
\label{Sec::Method}
Synchronization is a compulsory task in any communication system to correctly recover data from the bit stream. Although every communication transmitter and receiver has its clock source, its stability depends on multiple factors, such as the oscillator's quality, stabilization of the environment, and the underlying physics. 
%Hence, it implies that clock synchronization is in fact a compulsory task. 
This implies the necessity of clock synchronization.
One way to achieve clock synchronization is by sending a part, i.e., a decimated version, of the transmitter's clock via an optical channel\cite{Liu:10}. A second possibility is to lock the transmitter and the receiver to an external clock reference\cite{PhysRevA.92.052339,PhysRevA.91.042320}. However, clock recovery from  data is highly beneficial as it eliminates the need for additional signals and extra components. 
In classical optical communication, the ability to amplify data before distortion helps to preserve the bit stream integrity\cite{Gardner}, while, for quantum signals, the intrinsic single-photon regime rules out the possibility of regenerating all the information in a high-loss channel. In this situation, the success of clock recovery highly depends on channel loss, source repetition rate, and detection efficiency, which determines the available data to process and retrieve the clock rate. Additionally, sources of timing uncertainties, such as the pulse width, and jitters associated with the detectors and Time-to-Digital converter (TDC) limit the recovery of the timing of the data stream.

In general, a synchronization method includes two steps: clock synchronization, the task of coordinating the frequency of two independent clocks, and time synchronization; in the latter, the temporal offset between transmitter and receiver is compensated after the clock synchronization.

Time Interval Error (TIE) is a key parameter in analyzing clock drift which determines the time variation of a signal from the reference one. TIE is measured between their zero-crossing points\cite{Bregni_clock,bellamy} and is often represented as a fraction of the clock cycle. If T($t$) is the time of measuring a signal, the Time Error function TE($t$), as the signal temporal error with respect to its expected one $\text{T}_{\text{ref}}(t)$, is defined\cite{Bregni_clock}
\begin{equation*}
    \text{TE}(t) = \text{T}(t) - \text{T}_{\text{ref}}(t),
\end{equation*}
then for an interval $\tau$, TIE  is given by\cite{Bregni_clock}
\begin{equation*}
\begin{split}
    \text{TIE}_t(\tau) & = [\text{T}(t+\tau) - \text{T}_{\text{ref}}(t+\tau)] - [\text{T}(t) - \text{T}_{\text{ref}}(t)] \\
    & = \text{TE}(t+\tau) - \text{TE}(t).
\end{split}
\end{equation*}
TIE quantifies the variation of temporal error TE over a period of $\tau$.
The necessary condition  for the correction is given by:
\begin{equation}
    |\text{TIE}_t(\tau)| < \frac{\text{T}_{\text{Rx}}}{2},
    \label{Eq::CorrectionCriteria}
\end{equation}
where $\text{T}_{\text{Rx}}$ is the period, i.e., the inverse of the frequency, in which the receiver measures the consecutive pulses. Eq. \ref{Eq::CorrectionCriteria} implies that the accumulated error over time due to frequency mismatch does not shift the transmitted frame of data by one unit, leading to an uncorrelated transmitted and received sequence.
To evaluate TIE between the transmitter and receiver, we performed an \textit{independent clocks configuration}\cite{ITUT_Clcok} test where we replaced the transmitter's internal clock with a highly stable reference clock generated by an Agilent E8267D vector signal Generator.
This shows the isolated stability of the receiver clock which be can be observed in the dashed trace of figure \ref{Fig::clock_drift_all} as the drift in demodulation frequency.
%The performance of this clock replacement on the TIE can be seen as the demodulation frequency drift in figure \ref{Fig::clock_drift_all} as the line corresponding to 650000 detection per second. 
%In figure \ref{Fig::clock_drift_all} demodulation frequency corresponding to 650000 detection rate, 
%the impact of the TE sampling period on TIE measurement for various $T_{int}$ is shown. 
%a comparison of frequency drifts when the clock source is replaced w.r.t. internal transmitter's clock is represented.
The timing error is measured between the time-to-digital converter (TDC) internal clock and the Field Programmable Gate Arrays (FPGA) that provides the transmitter clock where we replaced one of the clocks with the signal generator.
The detection time of a signal varies with respect to the transmitter by a constant offset indicating the time-of-flight of the signal from the transmitter to the receiver. Furthermore, two additional changes over time, namely, frequency shift and frequency drift, affect the signal. The shift and drift lead to a frequency of detection $f_{\text{Rx}}=1/\text{T}_{\text{Rx}}$ that varies in time with respect to the transmitter frequency $f_{\text{Tx}} = 1/\text{T}_{\text{Tx}}$. 
%\begin{figure}
%    \centering
%    \includegraphics[width=0.49\textwidth]{TIE.png}
%    \caption{\textcolor{red}{I say we remove this figure but if it is to be kept, make it as is written in the text. Present the TIE as a fraction of clock cycle. This way it shows how far we got in terms of error from the expected one}. Time interval error measured between the transmitter clock and the receiver TDC clock for $\text{T}_{\text{int}}=5$ ms. }%Note that the TIE  measured w.r.t. the receiver frequency will be smaller.}
%    \label{Fig::TIE}
%\end{figure}

\begin{figure}
    \centering
    \includegraphics[width=0.49\textwidth]{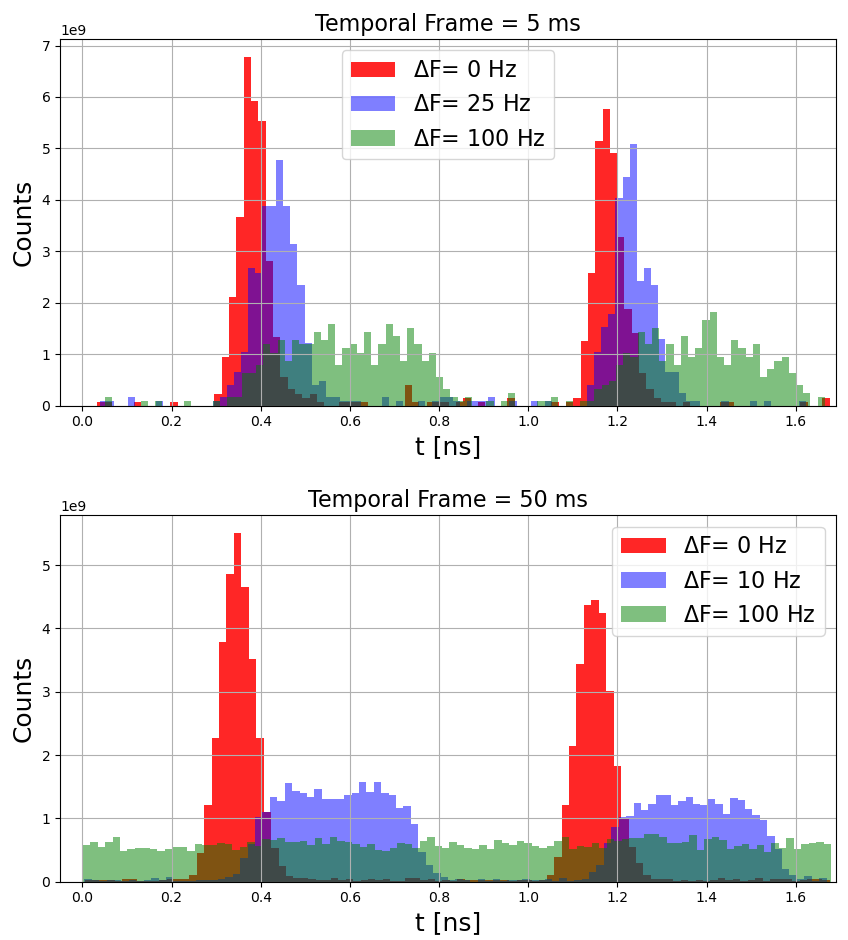}
    \caption{Representation of the effect of clock offset on the stream integrity for a frame of data acquired in 5 ms (top) and 50 ms (bottom) for different introduced frequency offsets obtained from our time-bin QKD source.}
    \label{Fig::ClockOffsetComparison}
\end{figure}

\begin{figure}
    \centering
    \includegraphics[width=0.48\textwidth]{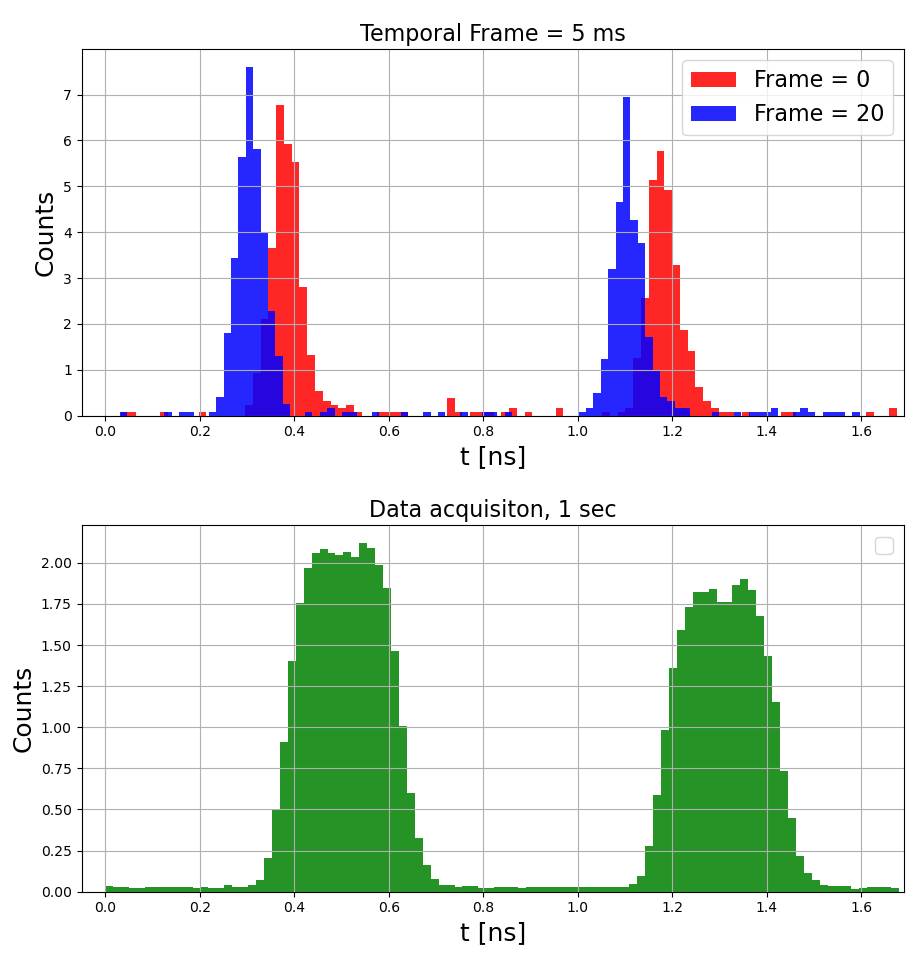}
    \caption{Data recovery represented in two steps. (Top) Data are divided into intervals of $\text{T}_{\text{int}}$. Each color demonstrates the temporal drift of consecutive frames of data. (Bottom) Drift compensated data.}
    \label{Fig::DriftvsNoDrift}
\end{figure}

While frequency shift - a constant difference between the transmitter's and receiver's clock rate - affects the registered time of signals by a fixed amount and maintains its periodicity, frequency drift - variation of the relative frequency of the source clocks over time - alters this feature and introduces TIE after clock recovery. Hence, in order to have a successful clock estimation by performing the FFT, it is necessary to find the largest interval that respects condition \ref{Eq::CorrectionCriteria}. If condition \ref{Eq::CorrectionCriteria} is not satisfied anymore for some $\tau$ inside a frame, some detections cannot be assigned correctly to their respective time bin anymore. In \cite{spiess2022clock}, a comparison of the stability of multiple clock sources is presented. 
In addition to condition \eqref{Eq::CorrectionCriteria}, computational cost puts a  barrier on the maximum length of the time frame in which one can perform FFT analysis. Lack of data sets a minimum duration time for a successful FFT result. On the other hand, a short time frame, as sample duration or observation time, influences the resolution in the frequency domain and can introduce an imprecision on the result of FFT. 
The limited resolution, especially at high repetition rates, affects the signal integrity severely, see Fig. \ref{Fig::ClockOffsetComparison} for a comparison of the clock imprecision on the histogram of acquired data.
To tackle this issue, \cite{PhysRevApplied.13.054041} proposes to find a first estimation by applying FFT analysis on a limited number of samples and later correct the result with the Least Trimmed Square method. The approach in \cite{spiess2022clock} is to perform a brute force around the first estimation to find the exact frequency. Here we adopt a similar technique and to reduce the computational cost, we apply the FFT in a short time frame. The resulting estimate is then used as the initial value for a fine correction of the frequency by measuring the drift introduced in time, see Figure \ref{Fig::DriftvsNoDrift}.
These analyses, however, assume a stable clock over the interval of interest. Note that in the absence of a low noise reference clock, the overall drift of the transmitter and receiver clock limits the duration of the acceptable time frame in which one can assume a constant clock offset.
If the duration of such a time frame is too short, such that no sufficient data is available, the clock recovery procedure will fail. We investigated the effect of data availability and clock drift in a simulation presented in the result section.

The four steps to recover the clock are as follows:
\begin{enumerate}

\item We divide the data into intervals (or frames) of $\text{T}_{\text{int}}$ and estimate the frequency by taking a fast Fourier transform of the data acquired in the first time interval with a sampling rate of at least twice the rate of the source. The data that is fed into the FFT is a binary array where each element corresponds to a sample, with a 1 for those time bins where a detection has been registered and a 0 for empty time bins. This procedure returns the term $f_{\text{Rx}}^0=1/\text{T}^0_{\text{Rx}}$ as the biggest coefficient of the FFT, the first estimate of the source repetition rate. 

\item We examine the clock drift over time. This knowledge allows us to estimate the number of following blocks for which the estimated $f_{\text{Rx}}$ is still valid. The clock drift is expected to be negligible in the observation intervals of interest in telecommunications\cite{Bregni_clock}, however, it highly depends on the clock stability.
Analyzing the histogram of the temporal distribution of data for consecutive blocks reveal a small varying temporal shift across them, see Fig. \ref{Fig::DriftvsNoDrift}a.

\item We measure the drift $\delta t$ in a time frame separated for $T$ in time w.r.t. the first one. Accordingly, the offset can be estimated by,
\begin{equation*}
    \delta F = \frac{\delta t}{T} f_{\text{Rx}}^0. % \frac{1}{f_{\text{Rx}}^0}.
\end{equation*}

\item We frequently check if the last frequency is still valid by examining the data and the accumulated drift in time. In general, clock drift determines how often the procedure needs to be repeated in order to correct the clock for the whole data set.
\end{enumerate}

Running the FFT method only for the first time frame drastically reduces the computational complexity. In step 3, we estimated the clock shift by analyzing the temporal drift in the first couple of frames. We then found the frequency offset, Fig. \ref{Fig::DriftvsNoDrift}-bottom and corrected the frequency by applying $\delta$F. Analyzing the data acquired in one acquisition time, cross-correlating them with the transmitted qubit sequence, and sifting, yields the final quantum bit error rate. However, if required, a higher precision can be achieved by applying an advanced optimization technique at the cost of a slight increase in computation complexity, introduced in  section \ref{subsec::driftTrack}.

\subsection{Clock drift tracking}
\label{subsec::driftTrack}
As visualized in Fig. \ref{Fig::ClockOffsetComparison}, even a small deviation in clock frequency causes the histograms to broaden significantly. However, the FFT method is only accurate up to $2/\text{T}_{\text{sample}}$. Given that according to the Nyquist theorem, the sampling rate must be twice the source repetition rate, high FFT resolution requires immense computational resources in analyzing high repetition rate sources.
Here we introduce a method to estimate the combined frequency drift of transmitter and receiver clock, which we call demodulation frequency drift, up to a few Hz. Additionally, this method can be used to track the drift of the required demodulation frequency, which corresponds to the combined drift of the clocks involved in the transmission and reception of the qubits.

% \begin{figure}[ht!]
%     \centering
%     \includegraphics[width=0.48\textwidth]{cost_funct.png}
%     \caption{Example for the cost function. At different offsets from the start of the data recording, the cost function is calculated on 10000 received photon time stamps, which corresponds to 9.7 ms in this case. The x axis displays the frequency offset from the demodulation frequency initially estimated using the Fourier transform method. The cost functions displays a clear minimum which corresponds to the optimal demodulation frequency. The minimum shifts with offset time, indicating a drift of the optimal demodulation frequency.}
%     \label{Fig::CostFunctionExample}
% \end{figure}

The approach relies on the fact that the histogram (as seen in Fig. \ref{Fig::ClockOffsetComparison}) will exhibit the narrowest peaks when the correct demodulation frequency $f_{Rx}^0$ is used for a given temporal frame. A cost function $\mathcal{J}$ that is minimal for the optimal frequency can be defined as follows:
\begin{enumerate}
    \item Take the histogram of the measured photon arrival times modulo $1/f^{0}_{Rx}$. This can be described by the discrete function
    \begin{align}
        % T &= \{ t \bmod (\frac{1}{f^{0}_{Rx}}) \mid t_{start} < t < t_{end} \}\\
        % \widetilde{t}_{j} \triangleq t_{j}\mod (\frac{1}{f^{0}_{Rx}}) \in T\\
        c_{i}(T, f^{0}_{Rx}) &:=  \sum_{t \in T}  \mathbf{1}_{\left[l_{i}, r_{i}\right]}(t)
    \end{align}
    where 
    \begin{align}
        T &= \{ t \bmod (\frac{1}{f^{0}_{Rx}}) \mid t_{start} < t < t_{end} \}
    \end{align}
    is a set containing all the arrival times of photons in the temporal frame given by $\left[t_{start}, t_{end}\right]$. We have the indicator function
    \begin{align}
        \mathbf{1}_{A}(t) :=     \begin{cases}
          1 & \text{if } t \in A\\
          0 & \text{if } t \notin A
        \end{cases}   
    \end{align}
    and $l_{i}$, $r_{i}$ are the bounds of the histogram bins $c_{i}$. Choosing reasonable histogram bin boundaries depends on the source repetition rate and the available time resolution. %, in our case, for the source described in section \ref{Sec::Experiment} 150 equally spaced bins performed best.
    %\sout{In practice, choosing 150 equally spaced bins works best.}
\item The cost function is defined as the sample variance of the histogram bin values:
    \begin{align}
        % \mu &= \frac{1}{N} \sum_j c_{j} \\
        \mathcal{J}(T, f^{0}_{Rx}) &= -\mathrm{Var}(c) = -\frac{1}{N} \sum_j (c_{j} - \mu)^{2}
    \end{align}
    where $\mu = \frac{1}{N} \sum_j c_{j}$ and $N = |T|$ is the number of elements in $T$.
\end{enumerate}

%The code snippet given below in Listing \ref{lst::cost_implementation} is an exemplary implementation of the cost function in python:

%\begin{lstlisting}[language=Python, caption=Cost function implementation in python, basicstyle=\footnotesize, label=lst::cost_implementation]
%import numpy as np
%def cost(
%    f_demod,
%    detection_times,
%    t_per_bin):
%    
%    fixed_bins = np.arange(
%        0,
%        2/f_demod,
%        t_per_bin
%                )
%    counts, _ = np.histogram(
%        detection_times % \
%        (2/f_demod),
%        bins=fixed_bins
%                )
%    return -np.std(counts)
%\end{lstlisting}
%The arguments of the function are \texttt{f\_demod} which is the demodulation frequency we want to calculate the cost function for, \texttt{detection\_times} which is an array containing the photon detection times and \texttt{t\_per\_bin} which is the width of each histogram bin.

%\large{\textcolor{red}{I am not sure about the code in the text, you decide if we want to keep it. I just thought it could be nice since this is less obfuscated by mathematical notation. Could also go into an appendix or Sup. material, I really dont care}}
%\normalsize

\begin{figure}[t]
   \centering
   \includegraphics[width=0.48\textwidth]{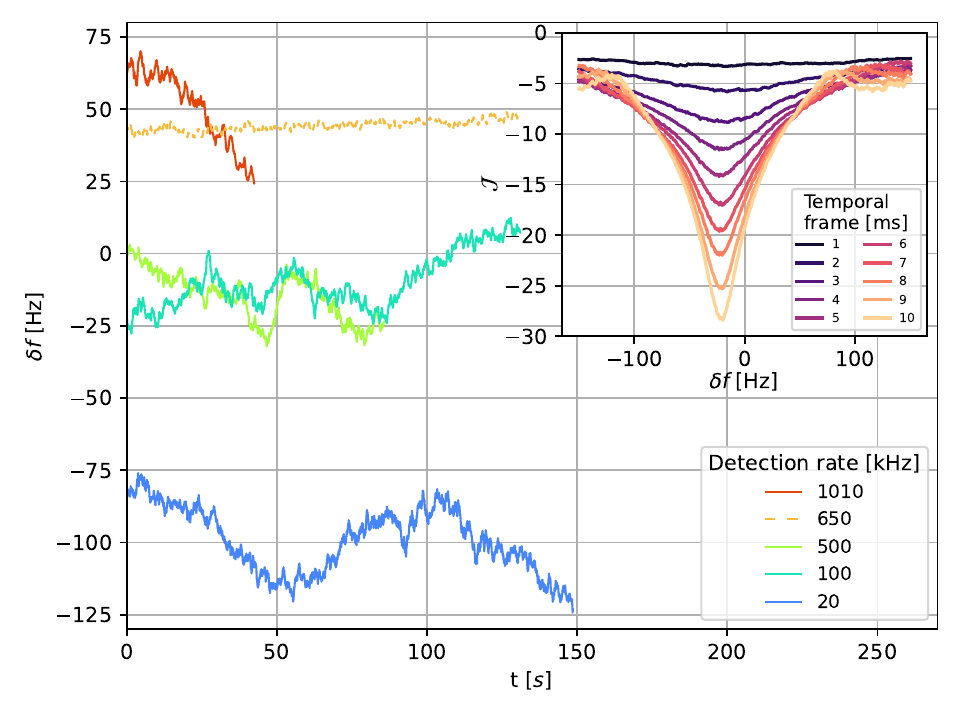}
   \caption{
   Drift of the optimal demodulation frequency over time for different detection rates, extracted from experimental traces. Using a sliding temporal frame approach of 30 ms with 20 ms overlap each, the cost function is optimized for every frame to extract the optimal demodulation frequency. For the dashed trace the transmission clock was locked to a stable reference clock, reducing the frequency drift over time significantly to only about 5 Hz over 130 s.
   The frequency $\delta f$ is given as the detuning from 1.1904734 GHz which is the FFT method estimate.
   The lengths of the traces differ due to different experimental measurement lengths.
   \textit{Inset}: Cost function $\mathcal{J}$ evaluated for different demodulation frequencies and different temporal frames. The detection rate in this exemplary photon arrival time data set is 100 kHz.
   The cost function displays a clear minimum which corresponds to the optimal demodulation frequency.}
   \label{Fig::clock_drift_all}
\end{figure}

The cost function is visualized in the inset of Fig. \ref{Fig::clock_drift_all} for a typical set of photon detection times. $\mathcal{J}$ is evaluated for different detunings $\delta f$ from the demodulation frequency that was first estimated using the FFT method. Further, the different traces show the cost function values for different lengths of temporal frames. For very short frames of only few ms, the function is flat and a minimum is hard to discriminate.
The detection rate, in such short time frames, determines the number of available data points to create a histogram and analyze it, resulting in a noisy cost function.
%In these short frames, dependent on the detection rate only few detections can be used to create the histogram, resulting in a noisy cost function. 
Using longer frames results in a statistically more significant histogram and the emergence of a clear minimum of the function. It is to note that for longer frames, local minima at large detunings arise. Therefore, a good starting point needs to be chosen to estimate the optimal detuning efficiently using a numerical optimizer.

By shifting the temporal frame along the recorded detection times in a sliding window approach and finding the optimal demodulation frequency for each offset, the drift in the demodulation frequency can be tracked.
For a sufficiently large temporal frame in which the noise on the cost function becomes negligible, we use the Nelder-Mead optimization algorithm \cite{gaoImplementingNelderMeadSimplex2012} to find the location of the minimum numerically. The required length of the frame depends on the detection rate with higher detection rates generally requiring shorter frames. Here we choose frames of 30 ms which overlap for 20 ms with the previous frame. The overlap improves optimizer convergence as the time difference between frames is smaller. The reduced time difference reduces frequency drift in between frames, therefore the starting point of the optimizer, which is determined by the result of the previous frame, is more likely to be close to the correct value and less likely to converge to a local minimum.
The resulting frequency drift traces are shown in Fig. \ref{Fig::clock_drift_all}.

The temporal behavior of the demodulation frequency is reminiscent of a random walk. The magnitude of the drift for the trace with a detection rate 650 kHz stands out as its fluctuation is lower. For this trace, the clock responsible for transmitting was referenced to a stable reference. This indicates that the phase noise of the transmitting clock is a significant contribution to the overall system drift.

This method can be useful to evaluate overall system performance and can track changes in performance when individual components are replaced.

\subsection{Clock drift simulation}
\label{subsec::driftSimu}
To investigate the effect of the clock frequency noise on the system stability a full simulation is created.
Frequency drift tracking analysis as in section \ref{subsec::driftTrack} reveals that the demodulation frequency drift can be approximated as a random walk process. While RF sources generally do not exhibit frequency noise that behaves like a random walk but have a more complex behavior, this approximation offers simplicity as the frequency noise behavior is controlled only by a single parameter. The magnitude of this frequency noise will be quantified in terms of full width at half maximum (FWHM) of the reference clock.

\begin{figure}[t!]
    \centering
    \includegraphics[width=0.48\textwidth]{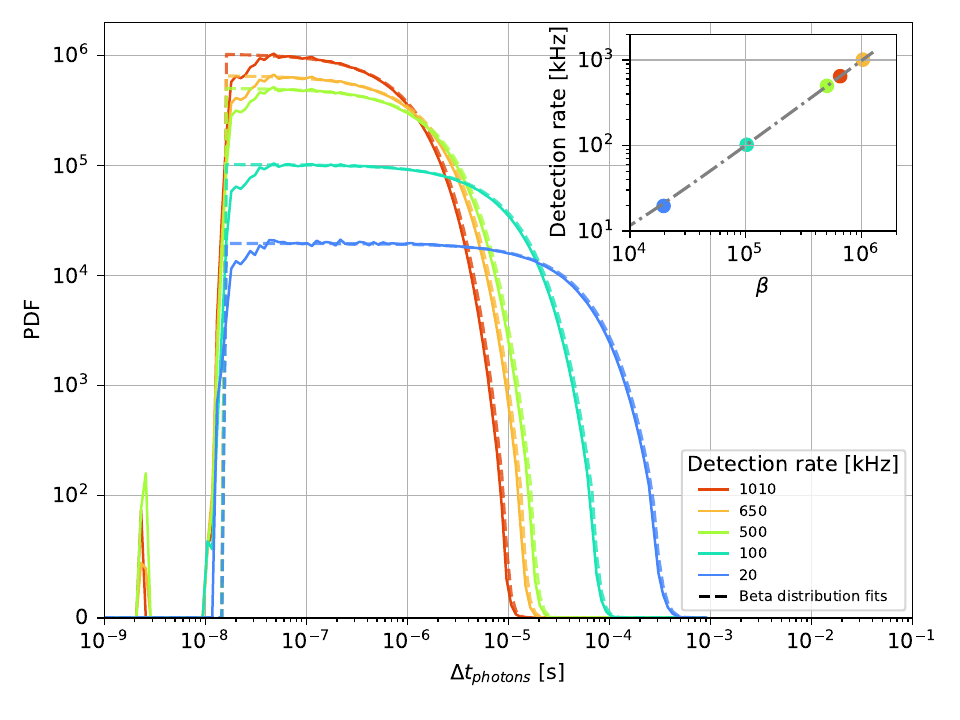}
    \caption{PDFs of the time difference of two successive detected photons from acquired data. Traces represent different data availability, i.e. varying photon detection rate. The corresponding dashed lines represent the PDFs of Beta distribution for which the parameters were fitted. The parameters $\alpha=1$, $x_{0}=15\cdot10^{-9}$ and $s=1$ are held constant while $\beta$ is optimized.
    Note that the y-axis uses symmetric logarithmic scaling with a linearity threshold of 100.
    \textit{Inset}: Linear correspondence of the fitted $\beta$ parameter of the distribution and the detection rate of the trace.
    }
    \label{Fig::beta_fit}
\end{figure}

Other relevant parameters in clock analysis are the source repetition rate and data availability. While available data determines the minimum temporal frame size for FFT analysis, the frequency noise magnitude determines the maximum frame size for a successful FFT before frequency drift makes the result uncertain. This simulation only considers a source repetition rate of 1.19 GHz in accordance with our experimental source.

Apart from the model for the clock frequency drift, a model for the lossy channel and photon detection is needed. We derive an empirical model by investigating the probability density distribution (PDF) of time differences between two successive photon detections from the acquired data. Fig. \ref{Fig::beta_fit} illustrates this distribution;
The detector has a dead time of approximately 30 ns, indicating that the time difference between two consecutive detections is dominantly above 30 ns. 
%\textcolor{red}{TODO: what is the actual dead time? From my fit it looks like the dead time is more like 10 - 15 ns}
This distribution can be fit to a scaled and shifted beta distribution \cite{Chattamvelli2021} which has the PDF
\begin{multline}
    f(x, \alpha, \beta, x_{0}, s) = \\
    \quad \frac{\Gamma(\alpha + \beta) (\frac{(x-x_{0})}{s})^{\alpha-1}(1-(\frac{(x-x_{0})}{s}))^{\beta-1}}{s \Gamma(\alpha) \Gamma(\beta)}
\end{multline}
where $\Gamma$ is the Gamma function.
The first two parameters $\alpha$ and $\beta$ define the shape of the distribution in the interval (0,1), while the last two parameters alter the location and scale of the distribution.
It turns out that 3 of the parameters can be held constant while only the $\beta$-parameter has to be tuned. In Fig. \ref{Fig::beta_fit} the PDFs estimated from experimental traces are shown next to the fitted beta distributions. The data is acquired from our time-bin setup, details in section \ref{Sec::Experiment}.

A linear correspondence can be extracted by investigating the fitted $\beta$-parameter against the different detection rates, as shown in the inset of Fig. \ref{Fig::beta_fit}.
This proportionality can be used to generate distribution parameters for any given detection rate, making the simulation highly flexible. By drawing random samples from the generated PDF we can simulate a series of photon detection times for a given detection rate.

% The relevant parameters in clock analysis are the source repetition rate, linewidth of the phase noise, and available data to work on. While available data determines the minimum frame size for FFT analysis, the linewidth of phase noise determines the maximum frame size for a successful FFT before frequency drift makes the result uncertain.
To complete the discussion we evaluate the impact of clock drift together with the size of available data. To quantify the impact of mentioned parameters on the qubit error rate when no measures are taken to correct for clock drift, we simulate the full QKD system for our fixed source frequency. First, for a given detection rate a few seconds of time tags are generated using the corresponding beta distribution.
%secondly, each timetag is time aligned such that each one is perfectly centered in the 800 ps time slots, which then represents timetags without noise as the ground truth. Now jitter is applied to each time tag according to the random walk noise model. Additionally to the random walk noise, also white noise simulating the detector uncertainty is added to each tag.
Secondly, the tags are temporally aligned such that they represent a perfect no-noise source. Then, clock frequency noise in the form of timing jitters is applied to each time tag according to the random walk noise model with a given noise magnitude. In addition to the random walk noise, the white noise simulating the detector uncertainty and optical pulse width is added to each tag. The zero-mean Gaussian generating the white noise is chosen to have a fixed experimentally determined value for the standard deviation of $\sigma=40$ ps.

\begin{figure}[ht!]
    \centering
    \includegraphics[width=0.48\textwidth]{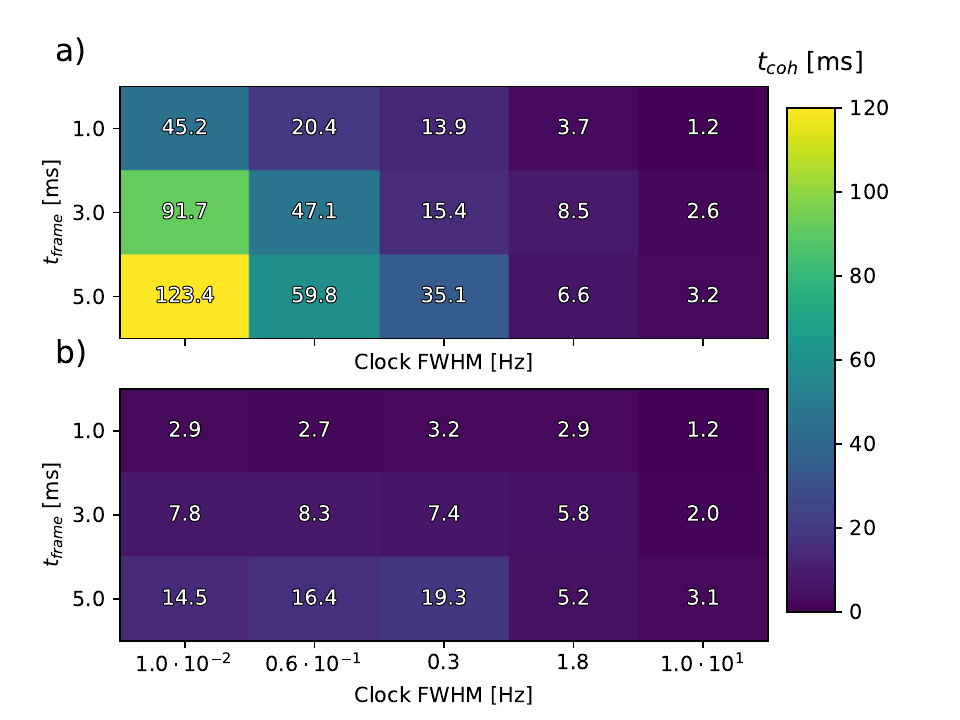}
    \caption{Median ``coherence time'' of the setup for a range of noise magnitudes and temporal frames. The median coherence time is defined as a measure of the lifetime of the system when no clock synchronization methods are applied. \textit{a)} shows the median lifetime when the demodulation frequency is estimated using the optimization method discussed in section \ref{subsec::driftTrack} while \textit{b)} shows the median lifetime for demodulation using only the FFT method. 
    A general trend of an increased lifetime with lower noise magnitude and higher estimation temporal frame can be observed. Additionally, the advantage of using the optimization method over using only the FFT method becomes clear.
    The detection rate used is 500 kHz and each point is averaged over 200 simulation runs.}
    \label{Fig::coherence_time}
\end{figure}

To start the frequency estimation, the described estimation method using the Fourier transform followed by the optimizer is employed on the predefined time frames. The extracted frequency is then used to demodulate the rest of the time tags and extract the noisy bit sequence. Comparing the noisy sequence to the true one allows us to calculate the QBER.
We define the ``coherence time'' of the system as the time it takes on average for the clock to drift so much that the QBER increases above 11\%, the threshold of BB84, due to loss of synchronization.
Figure \ref{Fig::coherence_time} shows the result of the simulation for our fixed source repetition rate and various clock-noise magnitudes and temporal frame lengths. It should be noted that a different source frequency will affect the result in two ways. Firstly, the available data at the same channel loss and the detector's efficiency would be different. Secondly, the sampling rate of FFT should be adjusted to the source frequency.

\section{Setup}
\label{Sec::Experiment}

\begin{figure*}[t!]
    \centering
    \includegraphics[width=0.98\textwidth]{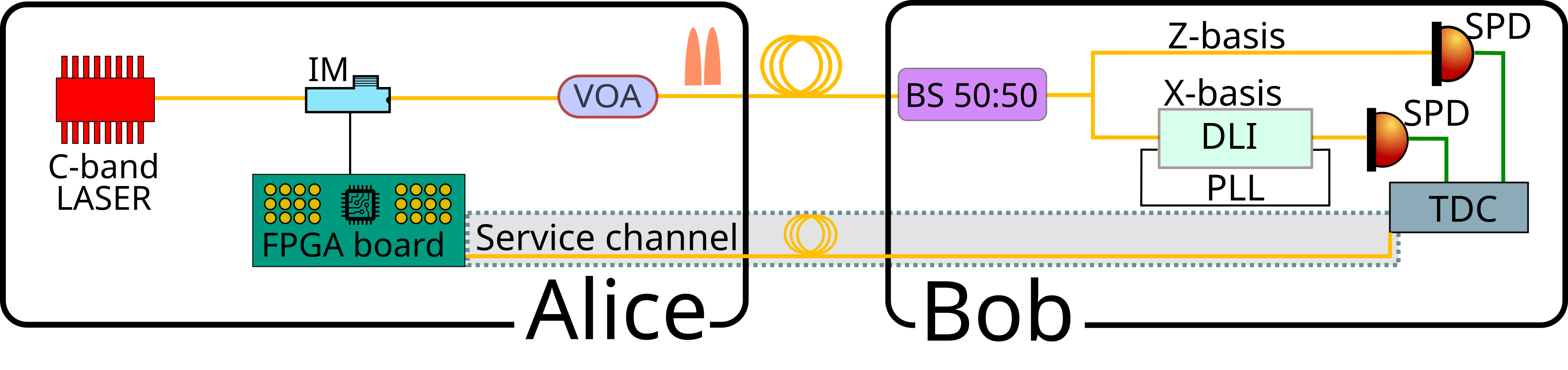}
    \caption{\textbf{Scheme of the setup.} Alice produces the quantum states by carving a continuous wave C-band laser. The carving is made utilizing an intensity modulator (IM) controlled by a Field Programmable Gate Array (FPGA board). Variable optical attenuation (VOA) decreases the optical power down to single-photon level. In the standard configuration, a stronger classical signal is produced by the FPGA and sent through a separate optical fiber or in a wavelength division multiplexing (WDM) approach (service channel), acting as a synchronization signal at 145 kHz. At Bob's side, a 50:50 beam splitter (BS 50:50) sends half of the states to the $\mathbf{Z}$-basis single-photon detector (SPD) and the other half to a delay line interferometer (DLI), stabilized by a phase lock loop (PLL), necessary for the $\mathbf{X}$-basis measurement. The events detected by the SPDs, together with the decoded synchronization signal, are finally sent to a time-to-digital converter and analyzed. Performing the clock recovery method, the entire service channel can be removed.}
    \label{Fig::setup}
\end{figure*}

We test the clock recovery procedure and its performance in a three-state BB84 QKD with one decoy state, implemented in time-bin encoding\cite{boaron_2018,molotkov_nazin_1996,shi2000quantum,fung2006security,Rusca2018_SecProofSimpleBB84,Rusca2018_FiniteKeyAnalysis,florence_field_trial}. In this protocol, the time of arrival of the photons within the frame of the time-bins is used to distinguish between the two states of the computational basis Z. Diagonal basis X, on the other hand, is not employed to extract a key but it is used for the security check; here the bit is characterized by a phase measurement on the two qubit's constituent pulses.
At the transmitter, time-bin qubits are generated by carving a continuous wave (CW) laser into the $\mathbf{Z}$ and $\mathbf{X}$ bases states. An FPGA drives an intensity modulator according to a pre-coded sequence to achieve a train of pulses with ~100 ps FWHM and close to 1.19 GHz repetition rate. Finally, a variable optical attenuator brings down the intensity to below the single-photon level per pulse. Decimating the qubit generation rate by 4095, a signal at 145.358 KHz is generated for synchronization purposes. However, this signal is used to compare the result of the clock recovery procedure in terms of performance and stability with the clocked QKD system.

At the receiver, a beam splitter implements a passive basis choice. The photons measured in the $\mathbf{Z}$ basis are directly sent to a superconducting nanowire single-photon detector (SNSPD), while in the $\mathbf{X}$ basis, an unbalanced Mach-Zehnder interferometer performs the phase measurement. The SNSPDs show 30 ns of dead time, around 100 Hz dark counts, less than 35 ps of timing jitter, and 83\% quantum efficiency. 
A TDC registers the detection events with a timing resolution of 1 ps. In normal operation mode, the TDC is locked to the synchronization or reference clock.

\section{Results}
\label{Sec::Result}
We performed the clock recovery procedure on our time-bin QKD system, featuring a 595 MHz qubit generation rate, for different channel loss values. In this demonstration, the sequence is a repetitive sequence of $2^{12}$ states randomly chosen from the Z and X bases. In a real scenario, however, a true random sequence produced by a quantum random number generator such as\cite{regazzoni2021high,Zahidy_2021_QRNG} is expected in order to guarantee security. We would like to stress that a time-bin source in effect shows a frequency twice its qubit generation rate since it has two pulses in each qubit. Furthermore, if the temporal distance between consecutive early and late pulses is not equal, it hinders the success of FFT in finding the precise $\text{T}_{\text{Rx}}$. This is also the case for us where the temporal distance of the consecutive early and late bins are 800 ps and 880 ps, respectively. 
%Figure \ref{Fig::HistoCompareClockedUnclocked} demonstrates the histogram of data obtained with $\text{T}_{\text{Tx}}$ and $\text{T}_{\text{Rx}}$. 

%\begin{figure}[h]
%    \centering
%    \includegraphics[width=0.47\textwidth]{Histo_Clocked_UnClocked_3.png}
%    \caption{Clock recovery output; histogram of data obtained with the expected clock rate (yellow) and recovered clock rate (Blue). These data were acquired at 22.5 dB of channel loss.}
%    \label{Fig::HistoCompareClockedUnclocked}
%\end{figure}

We chose $\text{T}_{\text{int}}=5$ ms as it respects the criteria \ref{Eq::CorrectionCriteria} and preserved the pulse width as expected from a clocked system. 
Furthermore, the performing FFT for $\text{T}_{\text{int}}=5$ ms with 400 ps sampling is not heavily time-consuming with a regular PC.
Since we perform FFT only on one $\text{T}_{\text{int}}$ of data, the overall process only slows down about 1.5 seconds the data processing of 1 second of data acquisition.

In order to test the performance of our method, we performed the 3-state BB84 with 50/50 states choice from Z and X bases and started from 22.5 dB of channel loss to prevent the saturation of our detectors. 
%It has to be noted that we started our acquisition from 22.5 dB of channel loss since we want to operate the systems outside the saturation regime of the detectors. 
We optimized, by mathematical simulation, the photons per pulse of signal and decoy to $\mu_1=0.7$ and $\mu_2=0.3$. Table \ref{Tab::QBERvsLoss} shows the QBERs achieved after clock recovery and sifting. In all cases, a temporal window of 200 ps was chosen to discard unwanted noise. Since we loop over a repetitive sequence of states, we didn't need to perform time synchronization and the offset between received and transmitted frames is compensated by performing a simple cross-correlation technique. However, time synchronization can be performed using only qubits with methods such as Qubit4Sync\cite{PhysRevApplied.13.054041} after clock recovery.
\begin{table}[h!]
\centering
\begin{tabular}{c | c | c } 
 %\hline
 Channel Loss  & Detection Rate  & QBER \\[0.5ex] 
 \hline
 22.5 & 1.49M & 1.19  \\ 
 %\hline
 28.5 & 359K & 1.16  \\
 %\hline
 34.5 &  94K & 1.26  \\
 %\hline
 40.5 &  13.6K & 3.74 \\
 %\hline
 42.5 &  6.2K & 6.77 
\end{tabular}
\caption{QBER and raw detection rate obtained from the raw data for $\mu_1=0.7$ after frequency recovery.}
\label{Tab::QBERvsLoss}
\end{table}

\begin{figure}[ht!]
    \centering
    \includegraphics[width=0.46\textwidth]{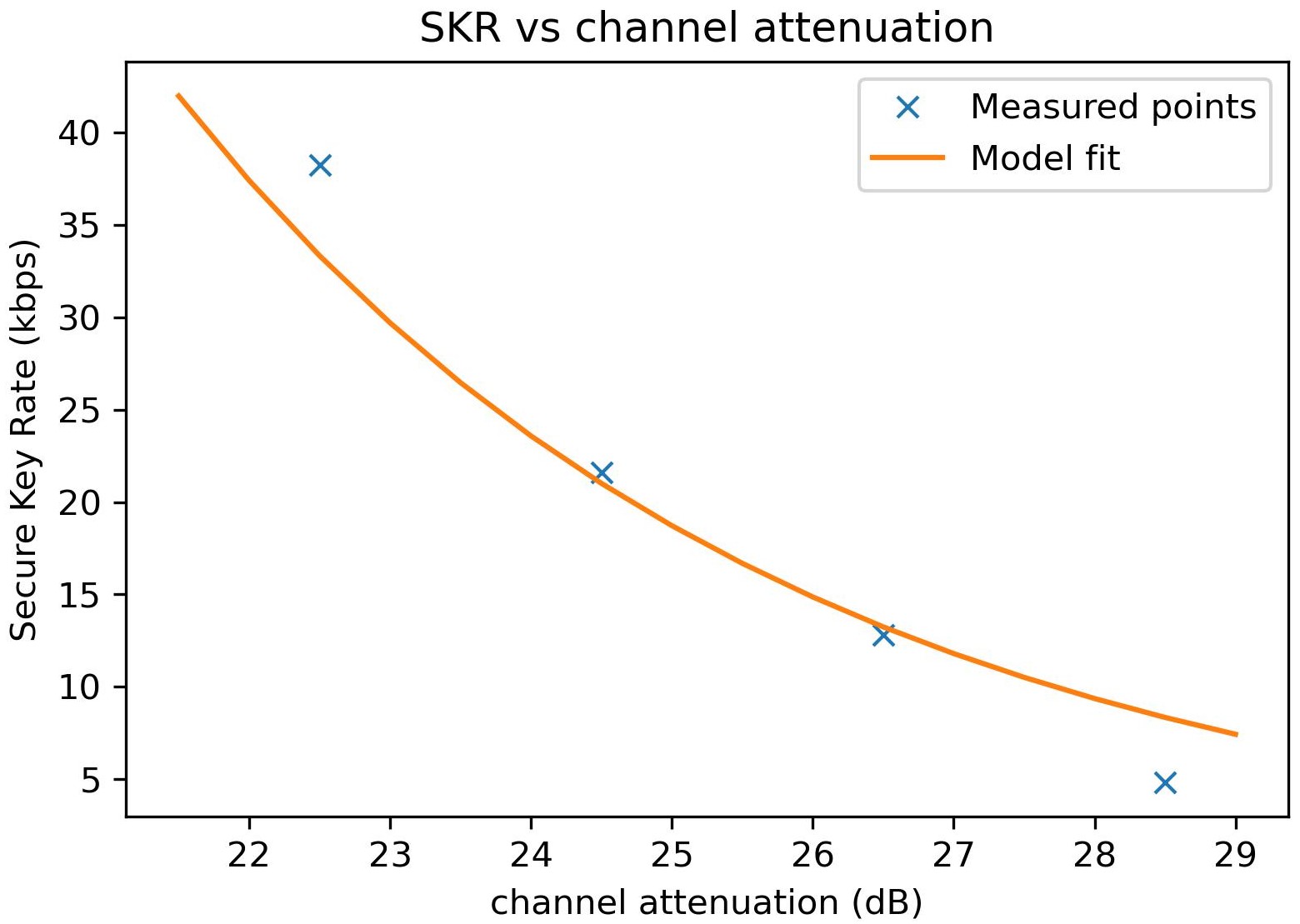}
    \caption{Secure key rate (SKR) vs. channel loss achieved for the QKD system without a shared clock. The obtained values (blue) fit the simulation model (orange line) for a system showing the same characteristics.}
    \label{Fig::LossSKR}
\end{figure}

We further analyzed the data for $\mu_2$ and estimated the SKR for security parameter $\epsilon_{sec}=10^{-12}$ according to the finite-size key analysis \cite{Rusca2018_FiniteKeyAnalysis}, Fig. \ref{Fig::LossSKR}. The clock recovery returned a QBER equal to 5.7\% for $\mu_2$ at 32.5 dB channel loss which made the extraction of secure keys impossible with the 3-states protocol due to lack of raw sifted data. %In case of a standard BB84 protocol the overall channel losses can be improved. 

%We would like to stress that the bottleneck in successful period recovery is not imposed by the FFT method and channel loss but rather was due to the failure of the algorithm to compensate for the relative drift of consecutive intervals. Fig. \ref{Fig::HistoCompare} shows the pulses of our time-bin qubits for different channel losses.
We observed that FFT works up to 48.5 dB of channel loss for our source and it returns the expected frequency. However, the noise prevents a successful optimization of the demodulation frequency. 

\section{Discussion}
\label{Sec:Discussion}

Summarizing, we presented a scalable and effective method for recovering the clock and the synchronization signal directly from the quantum states. The method has been tested in a real-time QKD system, based on time-bin encoding with one decoy-state method, and has demonstrated similar performances at moderate losses in terms of QBER and secret key rate compared to the same QKD system supported by the clock distribution feature. We further analyzed the performance of the approach by introducing a simulation technique based on the statistical features of general quantum communication. The simulation sets boundaries on the stability of the clock for a successful timing recovery.

The pulse generation rate in a QKD system sets a limitation on the performance and success of the method, especially at high losses. Small temporal separation requires us to perform FFT on considerably larger samples, while on the other hand $\text{T}_{\text{int}}$ should be chosen such that the condition of Eq.~\ref{Eq::CorrectionCriteria} is satisfied. $\text{T}_{\text{int}}$ also determines the broadening of the signal, which should be contained for better temporal filtering.

Further comparison of our method with other clock-recovery methods at a repetition rate close to the one of our system (about 1.2 GHz), would present a benchmark of clock recovery methods for QKD applications. Comparison with well-known methods such as Qubit4Sync\cite{PhysRevApplied.13.054041}, which exploits linear fitting, or with the one proposed in Ref.~\cite{QubitBasedSynch_Cochran} in which a Bayesian statistical approach is implemented, will be considered in future works.

%It is important to stress that since the frequency drift is measured to be constant at intervals of 1 sec, we expect the procedure to be successful even at higher loss values. 
We characterized the temporal drift of packets of data in each $\text{T}_{\text{int}}=5$ ms which shows an average of 190 ps. Using this value rough recombination of data at high losses is possible although at the price of broader pulses.  

Our demonstration paves the way toward an easier implementation of QKD systems into real telecommunication networks by removing a resource-consuming technique and replacing it with a cost-effective approach which could make the integration of QKD more attractive, and help the deployment of quantum technologies in real-life applications.

\section{Acknowledgements}
This work was partially supported by the Center of Excellence SPOC - Silicon Photonics for Optical Communications (ref DNRF123), the Innovationsfonden (No. 9090-00031B, FIRE-Q) the EraNET Cofund Initiatives QuantERA within the European Union’s Horizon 2020 research and innovation program (grant agreement No. 731473, project SQUARE), by the NATO Science for Peace and Security program (Grant No. G5485, project SEQUEL) and the programme Rita Levi Montalcini QOMUNE (PGR19GKW5T).

\section{Competing Interests}

The authors declare no competing financial or non-financial interests.

\section{Data Availability Statement}

The data that support the findings of this study are available from the corresponding author upon reasonable request.

\bibliography{bibliography}
\end{document}